\documentclass{article}
\usepackage{dcase2017,amsmath,graphicx,url,times,booktabs, tabularx}
\usepackage{amsmath,graphicx,amsfonts}
\usepackage{comment}
\usepackage{tikz}
\usetikzlibrary{matrix,shapes,arrows,positioning,chains, calc}

\title{DCASE 2017 Task 1: Acoustic Scene Classification Using Shift-Invariant Kernels and Random Features}

\name{Abelino Jim\'enez, $\;$ Benjam\'in Elizalde $\;$ and $\;$ Bhiksha Raj }\address{Carnegie Mellon University, Department of Electrical and Computer Engineering, Pittsburgh, PA, USA \\abjimenez@cmu.edu, $\;$ bmartin1@andrew.cmu.edu, $\;$ bhiksha@cs.cmu.edu}

\begin{document}

\ninept
\maketitle

\begin{sloppy}

\begin{abstract}
Acoustic scene recordings are represented by different types of handcrafted or Neural Network-derived features. These features, typically of thousands of dimensions, are classified in state of the art approaches using kernel machines, such as the Support Vector Machines (SVM). However, the complexity of training these methods increases with the dimensionality of these input features and the size of the dataset. A solution is to map the input features to a randomized lower-dimensional feature space. The resulting random features can approximate non-linear kernels with faster linear kernel computation. In this work, we computed a set of 6,553 input features and used them to compute random features to approximate three types of kernels, Gaussian, Laplacian and Cauchy. We compared their performance using an SVM in the context of the DCASE Task 1 - Acoustic Scene Classification. Experiments show that both, input and random features outperformed the DCASE baseline by an absolute 4\%. Moreover, the random features reduced the dimensionality of the input by more than three times with minimal loss of performance and by more than six times and still outperformed the baseline. Hence, random features could be employed by state of the art approaches to compute low-storage features and perform faster kernel computations.

\end{abstract}

\begin{keywords}
Acoustic Scene Classification, Laplacian Kernel, Gaussian Kernel, Cauchy Kernel, Kernel Machines, Random Features, DCASE Challenge
\end{keywords}

\section{Introduction and Related Work}

The DCASE Task 1 - Acoustic Scene Classification (ASC) aims to identify a recording as belonging to a predefined set of scene-classes that characterizes an environment, for example \textit{park}, \textit{home}, or \textit{office}. Typically, ASC approaches capture the diverse characteristics from the audio signal by computing different types of features, either hand-crafted~\cite{giannoulis2013detection,zhang2012semi,geiger2013large,metze2014improved,elizalde2016experiments} or derived from Neural Networks~\cite{zhang2017learning,arandjelovic2017look,aytar2016soundnet}. These features are commonly of high-dimensionality (up to ten of thousands) and state of the art ASC approaches classified them using Support Vector Machines, the best known member of kernel methods. 

Kernel methods have the kernel trick property, which employs a non-linear kernel function to operate in a high-dimensional space by computing the inner products between the all pairs of transformed input features. The inner products are computed and stored in the Kernel or Gram matrix, which computing time and storage complexity increases in the dimensionality and number of the input features. A solution is to compute random features~\cite{rahimi2008random}, which have been well studied mainly for shift-invariant kernels because of their closed form. The process maps the input features into a lower-dimensional random space. Then, the resulting random features approximate non-linear kernels with linear kernel computations, hence speeding up the kernel matrix generation.


In this paper, we evaluated our random features in the context the 2017 DCASE Task 1 - Acoustic Scene Classification~\cite{DCASE2017challenge}. First, we computed input features with over six thousand dimensions, then we computed random features to approximate three types of shift-invariant kernels, Gaussian, Laplacian and Cauchy. Both type of features, input and random, were classified using an SVM. Experiments show that the baseline is outperformed by 4\% by all features. Moreover, random features reduced their dimensionality by more than three times with minimal loss of performance and by six times and still outperformed the baseline. 

The paper is organized as follows: In Section~\ref{methods} we describe in detail the  kernel functions used. In Section 3 we present experiments and results for Task 1. Finally, in Section 4 we conclude discussing the scope of the presented technique as well as future directions.

\section{Methods: Shift-Invariant Kernels and Random Features}
\label{methods}

In this section we describe the computation of random features for three types of shift-invariant kernels in the context of SVM. Acoustic Scene Classification has been explored by state of the art approaches based on kernel methods, which find non-linear decision boundaries using a kernel function. The function takes input features (extracted from the audio) in a space $\mathcal{X}$ and yield output scene classes in $\mathcal{Y}$. In this paper, we consider $\mathcal{X} = \mathbb{R}^N$ and $\mathcal{Y}= \{1,2,...,C\}$. Moreover, the kernel function can be expressed as $K : \mathcal{X} \times \mathcal{X} \rightarrow \mathbb{R}$, which is positive-definite and yields the value corresponding to the inner product between $\phi({\bf x}_1)$ and $\phi({\bf x}_2)$. The function $\phi$ maps $\mathbb{R}^N$ to some space $\mathcal{H}$, which is generally of higher dimensionality and has better class separability.

However, computing the kernel function could become a prohibitive task if the dimensionality of the input, $N$, is large and if the size of the training set $n$ is large. This happens because in order to learn the decision boundary function $f$ from the input audio and the corresponding labels in the dataset $\{({\bf x}_1, y_1),({\bf x}_2,y_2),...,({\bf x}_n,y_n)\}$, we need to compute the value $K({\bf x}_i, {\bf x}_j)$ for every element $i,j\in \{1,...,n\}$. 

Therefore, our solution for this problem are {\em random features}, which approximate a kernel function by finding a map $\Phi$ from $\mathbb{R}^N$ to a low-dimensional random space $\mathbb{R}^M$, such that

\begin{eqnarray}
K({\bf x}_1 \, , {\bf x}_2) &\approx & \langle \Phi({\bf x}_1) \, , \, \Phi({\bf x}_2) \rangle
\label{randomFea_approx}
\end{eqnarray}

Although different random features mappings have been proposed for different kernel functions \cite{randFeat1} \cite{randFeat2}, we focused on random features for {\em shift-invariant} kernels. We say that a kernel is shift-invariant if for any $ {\bf x}_1, \, {\bf x}_2, {\bf z} \in \mathbb{R}^N$

\begin{eqnarray}
K({\bf x}_1 + {\bf z}\, , \, {\bf x}_2 + {\bf z}) &=& K({\bf x}_1 \, , \, {\bf x}_2)
\end{eqnarray}

Which is equivalent to say that, for any ${\bf x}_1$ and ${\bf x}_2$

\begin{eqnarray}
K({\bf x}_1 \, , \, {\bf x}_2 ) & = & K({\bf x}_1 - {\bf x}_2 \, , \, {\bf 0})
\end{eqnarray}

Shift-invariant kernels have been proven to admit a closed form of computing random features as stated by the use of the Bochner's theorem~\cite{rahimi2008random}. The function to compute random features $\Phi \, : \mathbb{R}^N \rightarrow \mathbb{R}^M$ is given by

\begin{eqnarray}
\Phi({\bf x}) = \sqrt{\frac{2}{M}} \cos \left( {\bf W} {\bf x} + {\bf b} \right)
\end{eqnarray}

where ${\bf W}$ is a $M \times N$ matrix, ${\bf b}$ is a vector with $M$ components and the $\cos$ function is element-wise. The randomness comes from the generation of the components of ${\bf W}$ and ${\bf b}$, where $b_i$ comes from a uniform distribution between $0$ and $2\pi$, and $w_{ij}$ comes from the Fourier Transform of the function $g(\delta) = K(\delta, {\bf 0})$. Therefore, the approximation stated in equation \ref{randomFea_approx}, depends on the kernel function involved and the distribution used to generate the matrix ${\bf W}$. 

In this paper, we focus on three well studied shift-invariant kernel functions, Gaussian, Laplacian and Cauchy. Their definition and corresponding distributions used to generate random features are described below.

\subsection{Gaussian Kernel and Random Features}
The Gaussian kernel, also known as Radial Basis Kernel, is perhaps the most popular after the Linear kernel. The Gaussian function employs the $\ell_2$ norm and we define,

\begin{eqnarray}
K({\bf x}_1, {\bf x}_2) & = & \exp \big( - \gamma \|{\bf x}_1 - {\bf x}_2 \|^2_2 \big)
\end{eqnarray}

To compute the random features, we generate the components of the matrix ${\bf W}$ according to a Gaussian distribution as follows,
$$w_{ij} \sim \mathcal{N}(0,2\gamma)$$

\subsection{Laplacian Kernel and Random Features}
The Laplacian kernel is similar to the Gaussian, but the main difference is that it employs the $\ell_1$ norm, where $\|{\bf x}\|_1 = \sum_{i=1}^N |x_i|$. In this work, we consider the Laplacian kernel,

\begin{eqnarray}
K({\bf x}_1, {\bf x}_2) & = & \exp \big( - \gamma \|{\bf x}_1 - {\bf x}_2 \|_1 \big)
\end{eqnarray}

To compute the random features, we generate the components of the matrix ${\bf W}$ according to a Cauchy distribution as follows,
$$w_{ij} \sim \text{Cauchy}(0,\gamma)$$

\subsection{Cauchy Kernel and Random Features}
The Cauchy kernel is less known in comparison to the previous two and computing this kernel can be even a more expensive task with high-dimensional vectors due to its mathematical form, hence benefiting more from the speed of processing random features. We define the kernel, 

\begin{eqnarray}
K({\bf x}_1, {\bf x}_2) & = &  \prod_{i=1}^N \frac{1}{1 + \gamma^2(x_{1i} - x_{2i})^2}
\end{eqnarray}

To compute the random features, we generate the components of the matrix ${\bf W}$ according to a Laplace distribution,
$$ w_{ij} \sim \text{Laplace}(0,\gamma)$$

\begin{table*}[!htbp]
\centering
\caption{The class-wise accuracy of the four different kernels outperformed the baseline of the development set. *Note that the linear kernel is without using the random features.}
\label{tab:task1}
\begin{tabular}{l|c |c |c |c|c}
Acoustic scene & Baseline & Linear* Kernel  & Gaussian Kernel & Laplacian Kernel & Cauchy Kernel \\
\hline
Beach              &	75.3 \% &  78.2 \% & 78.8 \% & 77.2 \% & 77.9 \% \\
Bus                &	71.8 \% &  93.3 \% & 93.6 \% & 92.0 \% & 92.3 \% \\ \hline
Cafe/Restaurant    &	57.7 \% &  79.2 \% & 76.9 \% & 82.7 \% & 78.5 \% \\
Car                &	97.1 \% &  95.2 \% & 94.9 \% & 94.2 \% & 95.5 \% \\
\hline
City center        &	90.7 \% &  92.0 \% & 91.0 \% & 92.3 \% & 89.4 \% \\
Forest path        &	79.5 \% &  87.8 \% & 89.1 \% & 85.9 \% & 87.2 \% \\
\hline
Grocery store      &	58.7 \% &  74.7 \% & 74.7 \% & 74.7 \% & 74.0 \% \\
Home               &	68.6 \% &  66.9 \% & 66.3 \% & 67.3 \% & 66.3 \% \\
\hline
Library            &	57.1 \% &  66.0 \% & 65.7 \% & 58.3 \% & 65.1 \% \\
Metro station      &	91.7 \% &  81.4 \% & 82.7 \% & 83.7 \% & 83.3 \% \\
\hline
Office             &	99.7 \% &  90.4 \% & 89.7 \% & 92.9 \% & 90.4 \% \\
Park               &	70.2 \% &  62.2 \% & 65.1 \% & 61.5 \% & 60.9 \% \\
\hline
Residential area   &	64.1 \% &  62.2 \% & 65.7 \% & 68.3 \% & 63.5 \% \\
Train              &	58.0 \% &  59.0 \% & 57.7 \% & 65.7 \% & 61.9 \% \\
Tram               &	81.7 \% &  81.1 \% & 82.7 \% & 84.3 \% & 81.7 \% \\
\hline
\textbf{Overall}	& \textbf{74.8 \%}  & \textbf{78.0 \%} & \textbf{78.3 \%} & \textbf{78.8 \%} & \textbf{77.9 \%}\\
\end{tabular}
\end{table*}

\subsection{Training SVMs with Random Features} 

An SVM is a kernel method that can perform non-linear classification by solving the quadratic optimization of the dual form and taking advantage of the kernel trick~\cite{Bhishop}. The kernel trick uses a non-linear function to map the input features into a high-dimensional feature space by computing the kernel matrix. 

An SVM using a non-linear shift-invariant kernel using the input features could be approximated by a linear SVM using the random features. The kernel matrix resulting from computing the inner product between the random features correspond to an approximation of the kernel matrix using the input features and the shift-invariant kernel. The linear computation has an important implication because there are libraries optimized for these problems.

\section{Experimental Setup and Results}

Our two set of experiments addressed the DCASE Task 1 - Acoustic Scene Classification~\cite{DCASE2017challenge}. We evaluate and compare the performance of the input features using SVMs with three non-linear shift-invariant kernels against the random features corresponding to the three kernel types using linear SVMs. Both pipelines are illustrated in Figure~\ref{figs:diagram}.

\begin{figure}[h]
   \centering
     \includegraphics[width=0.5\textwidth]{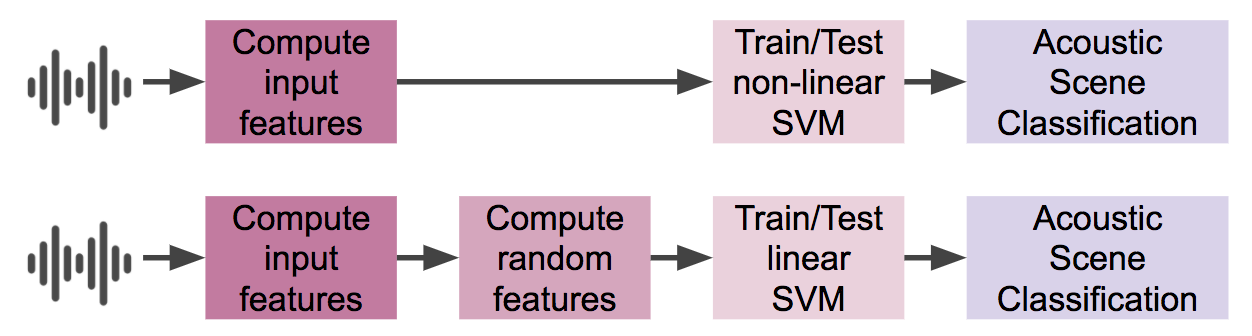}
     \caption{The acoustic scene dataset is used to extract input features for each recording. Then, the input features are used to train the SVM in two different ways. One is to pass the features directly to a non-linear shift-invariant kernel SVM, and the other is to first compute the random features and then pass them to a linear kernel SVM. Lastly, the trained SVM is used for multi-class classification on the test recordings.}
     \label{figs:diagram}
\vspace{-0.1in}
\end{figure}

\subsection{Acoustic Scene Dataset}
For our experiments we used the development set of the ``DCASE: TUT Acoustic Scenes 2017'' dataset~\cite{Mesaros2016_EUSIPCO}. It consists of recordings from various acoustic scenes of 3-5 minutes long divided into 4 cross-validation folds. The original recordings were then split into segments with a length of 10 seconds. Recordings were made using a binaural microphone and a recorder using 44.1 kHz sampling rate and 24 bit resolution. The 15 acoustic scenes are: \textit{Bus, Cafe / Restaurant, Car, City center, Forest path, Grocery store, Home, Lakeside beach, Library, Metro station, Office, Residential area, Train, Tram, Urban park.}

\subsection{Compute Input Features}
 
We extracted a large set of audio features proposed in~\cite{geiger2013large}, which are later used to compute the random features. The set include different features to capture different information from the acoustic scenes, which consist of multiple sound sources. The set is computed with the open-source feature extraction toolkit openSMILE~\cite{eyben2010opensmile} using the configuration file \textit{emolarge.conf}. The features are divided in four categories: cepstral, spectral, energy related and voicing and are extracted every 10 ms from 25 ms frames. Moreover, included are functionals, such as mean, standard deviation, percentiles and quartiles, linear regression functionals, or local minima/maxima. The total dimensionality of the feature vector is 6,553. 

\subsection{Input Features and Non-linear SVM} 
The first set of experiments aimed to evaluate our large set of input features and non-linear SVMs in ASC. We used the input features to train the three types of non-linear shift-invariant SVMs, also, we included the linear kernel (without random features). The SVM parameter $C$ was tuned using a search grid on the linear kernel and was fixed in all cases to $C=100$ and the performance was measured using accuracy. The accuracy is the average classification accuracy over the 4 validation folds provided for this challenge. Additionally, we explored different values for $\gamma$, obtaining the best results with $\gamma = 2^{-18}$ for Gaussian Kernel, $\gamma = 2^{-14}$ for Laplacian Kernel, and $\gamma = 2^{-8}$ for Cauchy Kernel. Before training the models, in each fold we normalized the input features with respect to the training set. We computed the mean and the standard deviation using each feature file and then subtracted the mean and divided by the standard deviation every file in the training and the testing sets. 

The classification performance for all kernel types was similar as shown in Table~\ref{tab:task1}. Generally, non-linear kernels tend to perform better than linear kernels for ASC~\cite{giannoulis2013detection}. However, it's not uncommon to have a similar performance if the class separability given by the features is not so complex, which could be our case. Among our best classified scene-classes we have \textit{Bus}, \textit{Cafe/Restaurant} and \textit{Grocery store} with improvements of up to 25\%.  

\subsection{Random Features and Linear SVM}
The second set of experiments aimed to show that the use of random features and linear SVM have a similar performance to the non-linear SVMs. For this, we used the training and testing input features to compute the random features corresponding to each of the three shift-invariant kernels described in Section~\ref{methods}. Then, these random features were used to train the SVM with linear kernel.

The performance of employing the random features indeed compared to the one of the input features with non-linear SVM as shown in Table \ref{table2}. We can see that the results improve as $M$, the dimensionality of the random features increases, hence showing minimal loss of performance compared to the previous non-linear SVMs. Notice that $M$ is always lower than the original dimensionality of our input features. If we would have further increased the value of $M$, we would have an improvement of performance until convergence to the values from Table~\ref{tab:task1}.

\begin{table}
\caption{Overall accuracy by computing random features and using a linear SVM depending on the value of $M$, which is the dimensionality of the random features. Note that all the $M$ values are smaller than the input features (6,553) and the larger the values the closer the get to the ones in Table~\ref{tab:task1}.} 

\begin{tabular}{ | p{0.5cm} | p{2.1cm} | p{2.2cm} | p{2.1cm} |}
    \hline
    M  & 
    Gaussian Kernel  $(\gamma = 2^{-18})$ & 
    Laplacian Kernel $(\gamma = 2^{-14})$ & 
    Cauchy Kernel $(\gamma = 2^{-8})$  \\ \hline \hline
$2^5$     &  $\quad$ 50.4 \%  &  $\quad$ 49.8 \%  & $\quad$  48.7    \% \\ \hline
$2^6$     &  $\quad$ 57.3 \%  &  $\quad$ 56.0 \%  & $\quad$  56.2    \% \\ \hline 
$2^7$     &  $\quad$ 64.4 \%  &  $\quad$ 61.5 \%  & $\quad$  62.9    \% \\ \hline 
$2^8$     &  $\quad$ 69.1 \%  &  $\quad$ 66.0 \%  & $\quad$  67.9    \% \\ \hline 
$2^9$     &  $\quad$ 73.0 \%  &  $\quad$ 67.2 \%  & $\quad$  72.7   \% \\ \hline 
$2^{10}$  &  $\quad$ 75.3 \%  &  $\quad$ 70.3 \%  & $\quad$  75.1   \%  \\ \hline 
$2^{11}$  &  $\quad$ 76.1 \%  &  $\quad$ 73.0 \%  & $\quad$  75.7   \%  \\ \hline 
$2^{12}$  &  $\quad$ 77.2 \%  &  $\quad$ 75.8 \%  & $\quad$  76.9   \%  \\ \hline 
    \end{tabular}
  
\label{table2}
\end{table}

\subsection{Acoustic Scene Classification} 

The reported DCASE baseline \footnote{http://www.cs.tut.fi/sgn/arg/dcase2017/challenge/task-acoustic-scene-classification} was tailored to a multi-class single label classification setup, with the network output layer consisting of softmax type neurons representing the 15 classes and frame-based decisions were combined using majority voting to obtain a single label per classified segment. The classification resulted in 74.8\% accuracy, which was outperformed by an absolute 4\% using the input features and the SVM with Laplacian Kernel.

In relation to random features, we can observed that already with a reduction of dimensionality of $M = 2^{10} = 1024$, we obtained a similar performance to the DCASE baseline (74.8\%) for the Gaussian (75.3\%) and the Cauchy (75.1\%) kernels. Thus, reducing the dimensionality up to one sixth from the original 6,553 dims. Moreover, with a reduction of dimensionality of $M = 2^{12} = 4096$, we obtained a minimal loss of an absolute 1\% for the Gaussian and Cauchy kernels. Note that for the DCASE challenge we submitted a system using the input features and the Laplacian kernel SVM. The overall classification was 60\% in comparison to the reported baseline of 61\%. 

The advantage of random features is that they can reduce significantly the amount of the storage and the computational processing by reducing the dimensionality and using linear inner products. Unlike other dimensionality reduction methods, such as PCA, the technique presented in this paper does not need heavy computation cost, like computing eigenvectors, but we just need to generate random numbers with the appropriate kernel-related distribution. Moreover, other machine learning algorithms that employ kernels could be benefited.

Multiple applications can take advantage of random features. For example, state of the art techniques are currently dealing with features of over 10,000 dims and with hundreds of thousands of segments~\cite{zhang2017learning,arandjelovic2017look,aytar2016soundnet}, which are then passed to linear SVMs. Another example is when the audio is recorded on local devices and sent to the cloud, this technique helps to compress information by reducing the cost of transmission and preserve privacy. For instance, we can compute the random features keeping the parameters $W$ and $b$ private. Thus, we can still process the transformed data in the cloud with linear models without revealing the actual data.
    
\section{Conclusions}
In this paper we have addressed Task 1 - Acoustic Scene Classification and have outperformed the baseline accuracy by 4\% using a large set of acoustic features and non-linear SVMs. Additionally, we computed random features that approximated three types of shift-invariant kernels, which were passed to a linear SVM. We showed how the dimensionality can be decreased by one sixth with a minimal degradation of performance of about 1\%. The results may have significant implications in the big data context, where high dimensional features must be stored and quickly processed. 

\bibliographystyle{IEEEtran}
\bibliography{refs}
%
%
%
%
%
%
%
%
%

\end{sloppy}
\end{document}